\documentclass[11pt]{article}
\usepackage[font=small,labelfont=bf]{caption}
\usepackage{listings}
\usepackage{subcaption}
\usepackage{amsmath,amssymb}
\usepackage[left=1in, right=1in, top=1in, bottom=1in]{geometry}
\usepackage{verbatim}
\usepackage{titling}
\usepackage{parskip}

\usepackage{todonotes} 

\usepackage{appendix}

\usepackage[english]{babel}
\usepackage[utf8]{inputenc}
\usepackage{slashed}
\usepackage{amsfonts}
\usepackage{amsthm}
\usepackage{amsmath,amssymb}
\usepackage{empheq}
\usepackage{dsfont}
\usepackage{xcolor}
\usepackage{hyperref}

\newcommand{\braket}[1]{\left\langle \right. #1 \left. \right\rangle}

\newcommand{\ER}{Erd\H{o}s-R\'enyi }

\usepackage{xcolor}

\usepackage{soul}

\title{Network Inference from Population-Level Observation of Epidemics}
\author{F. Di Lauro$^1$, J.-C. Croix$^1$, M. Dashti$^1$, L. Berthouze$^2$, I.Z. Kiss$^1$}
\date{$^1$ Department of Mathematics, University of Sussex, Falmer, Brighton BN1 9QH, UK\\
	  $^2 $ Centre for Computational Neuroscience and Robotics, University of Sussex, Falmer BN1 9QH, UK\\ 
	  \vspace{1cm}\today
	  }

\begin{document}
\maketitle
\begin{abstract}
Using the continuous-time susceptible-infected-susceptible (SIS) model on networks, 
we investigate the problem of inferring the class of the underlying network when
epidemic data is only available at population-level (i.e. the number of infected individuals at a finite set of discrete times of a single realisation of the epidemic), 
the only information likely to be available in real world settings.
To tackle this, epidemics on networks are approximated by a Birth-and-Death process which keeps track of the number of infected nodes at population level. 
The rates of this surrogate model encode both the structure of the underlying network and disease dynamics. We use extensive simulations over Regular, Erd\H{o}s-R\'enyi and Barab\'asi-Albert networks to build network class-specific priors for these rates.
We then use Bayesian model selection to recover the most likely underlying network class, based only on
a single realisation of the epidemic. We show that the proposed methodology yields good results on both synthetic and real-world networks.

\end{abstract}


\section{Introduction}

Networks are an important tool for modelling systems with many interacting parts such as epidemics spreading within a population or neuronal activity in the brain. 
Indeed, the intricate interplay of many individual well-defined units can be captured by the links of a network, and this can be done with an unprecedented level of detail 
\cite{newman2003structure,keeling2005Interface,Danon2011,Porter2016,Kiss2017}. For instance, directed, weighted or temporal links can all be 
considered within this modelling paradigm. The power of this approach is particularly clear in the study of spreading processes, with epidemics on networks having been extensively studied as a diffusion process between nodes mediated by the links of the network. It is well established that the structure of the network has a 
profound impact on how diseases invade, spread and how to best control them. This is particularly well understood for degree heterogeneity and assortativity/disassortativity, and to a lesser extent, for clustering, the propensity of nodes that share a common neighbour to be connected \cite{pastor2014epidemic,Kiss2017}.

However, depending on the field of application, the precision to which the underlying network is known can vary greatly, from absolute (when full description is available) to absent (when a description is entirely lacking). For example, whereas some technological networks can be mapped out to a great degree of detail, social networks can be challenging to query~\cite{Brugere2018}. 
This has resulted in a significant amount of research aimed to develop methods for link prediction (for a survey, see \cite{Brugere2018}). 
Instead of assuming the availability of explicit information about nodes and edges, these methods rely on `observables' from 
dynamical processes taking place on the network, under the assumption that these provide latent information about the missing underlying network structure. 
In the framework of epidemics on networks this suggests that it is possible to get insights about the structure of the network by observing 
quantities of interest at node and perhaps population level. 
Indeed, the inverse problem of inferring networks from epidemic data has been the subject of great scrutiny. 

In particular, in the context of statistical inference, this task has been approached by either formulating it as a likelihood optimisation problem \cite{GomezRodriguez2010, Netrapalli2012, Myers2010, Du2012, Rodriguez2014} or using Bayesian inference \cite{Oneil1999,Britton2002,Groendyke2011,Grenfell2013, Dutta2018}. Compared to maximum likelihood optimisation methods (e.g. independent cascade model \cite{GomezRodriguez2010}), the Bayesian inference is usually based on a smaller number of observations of the epidemic \cite{Britton2002, Groendyke2011,  Dutta2018}. However, both network inference approaches (explicit link inference and inferring parameters of a known network model) lead to good estimates for the network and parameters of the epidemic dynamics. Moreover, there is an interesting tradeoff between them. The former is able to identify the adjacency matrix, but requires the observation of a large number of cascades, whereas the latter can only infer some structural parameters (such as the probability of a link between two nodes), but relies on fewer observations. Recently, it has been conjectured~\cite{prasse2018exact} that an exact (link-by-link) reconstruction of networks might not be feasible due to requiring a subexponentially increasing number of observations and an exponentially increasing computation time with respect to the number of nodes in the network.

A common feature of the above mentioned work is their reliance on the availability of detailed data at node level, such as the complete temporal knowledge of all cascade trees 
in~\cite{GomezRodriguez2010} or the observation of all the removal/infection times in the Bayesian framework of~\cite{Britton2002}. However, in most real-world scenarios, such detailed information is unlikely to be available. A more reasonable expectation is to be provided with population-level observations, that is, the number of infected nodes in the whole network at various times. 
As far as we are aware (for a survey, see~\cite{Vernon2011}), there is no research that specifically addresses the problem of network inference based purely on such kind of observations. Since inferring detailed network properties is not plausible, our aim in this paper is to establish the feasibility of inferring the class of the underlying network. We do so within the framework of continuous-time SIS epidemics on networks when only population-level  data from a single realisation of the epidemic are available. 

We treat this problem as an inverse problem and adopt a Bayesian approach which involves the following steps: 
\begin{enumerate}
\item[ (a)] propose a parametric forward model that reproduces network/population-level dynamics and reflects network structure; 
\item[(b)] build a prior distribution for these model parameters on a network class basis;
\item[(c)] use the posterior measure to identify the most likely network class.
\end{enumerate}

A complete description of the SIS dynamics on a network with $N$ nodes requires to solve $2^N$ equations, one per possible state. The distribution of population-level statistics in time can be described via the count of the number of infected nodes in this dynamics; however, this process scales exponentially with the size of the network. Here, we take a different route and choose to use a surrogate model to represent the evolution of the count of infected nodes in the population. A reasonable candidate for this is a Birth-and-Death process (BD), see~\cite{Nagy2014}, characterised by only $N+1$ equations and $2(N+1)$ free parameters, the rates of infection and recovery, that need to be tuned to best represent the exact model. Whilst the rates of recovery are network independent and known exactly, the rates of infection in the surrogate model are more challenging to define.

In this work, the rates of infection in the surrogate model are provided by a simple parametric model, together with an estimation procedure based on extensive and detailed simulations of epidemics on three classes of well-known random networks: Regular, Erd\H{o}s-R\'enyi and Barab\'asi-Albert. This procedure leads to distinct rate models for the three classes of networks. These observations are encapsulated in a prior distribution for the rates of the BD process.

Finally, when one observes a single epidemic through population-level data, our prior and forward model can be used within a Bayesian model selection framework to identify the most likely underlying network class. It is worth noting that this framework is versatile enough to be used in conjunction with any set of population-level epidemic data, as it will still output the most likely network class, that is, the closest class (in terms of heterogeneity of the degree distribution) to that of the true underlying network. 

The paper is structured as follows. In Section~\ref{sec:forward_model} we describe the BD surrogate forward model together with a three-parameter model for its rates of infection. Section~\ref{sec:bayesian} includes all aspects of the Bayesian approach we used, from building priors to model selection and model validation/stress testing. We conclude with a discussion and further research directions in Section~\ref{sec:discussion}.


\section{The forward model}
\label{sec:forward_model}

A population of $N$ individuals is considered with the contact structure between individuals described by an undirected network with adjacency matrix $G=(g_{ij})_{i,j=1,2,\dots, N}$ 
where $g_{ij}=1$ if nodes $i$ and $j$ are connected and zero otherwise. Self-loops are excluded, so $g_{ii}=0$ and $g_{ij}=g_{ji}$ for all $i,j=1,2, \dots N$. The standard SIS epidemic dynamics on a network is considered, which is driven by two type of events: (a) infection and (b) recovery from 
infection. Infection can spread from an infected and infectious node (I) to any of its susceptible neighbours (S) and this is modelled as a Poisson point process with per-link infection rate $\tau$. 
Infectious nodes recover at constant rate $\gamma$, independently of their neighbours and become susceptible again. The resulting model is a continuous-time Markov Chain over a state space with $2^N$ elements, consisting of all arrangements of length $N$ with each entry being either S or I. While this is easy to formalise and write down theoretically, the numerical integration of the system becomes intractable even for modest values of $N$~\cite{simon2011JMB,Danon2011,simon2013exact,Kiss2017}. This motivates us to use a surrogate model, offering sufficient flexibility to approximate the time evolution of the number of infectious nodes in the network.
%
%

\subsection{Birth-and-death approximation of SIS epidemics}
\label{subsec:BDapprox}

We use a BD process, a continuous-time Markov chain with state space $\{0,\dots,N\}$ and transitions of unit size, as the surrogate model. The up-jumps or infections are described by rates $a_k$, that is, the rates of infection in the presence of $k$ infected nodes and encode the network structure. The down-jumps or recoveries are described by rates $c_k=\gamma k$ and are network independent. Hence, the transition probabilities of the surrogate process are given by the following forward Kolmogorov (or Master) equation:
\begin{equation}
\forall k\in\lbrace 0,\dots,N\rbrace,\;\dot{p}_{k_0,k}(t) = a_{k-1} p_{k_0,k-1}(t) -  \left( a_k + c_k \right) p_{k_0,k}(t) + c_{k+1} p_{k_0,k+1}(t),
\label{eq:mastereq}
\end{equation}
together with $a_{-1}=c_{N+1}=0$ and an initial condition $k_0\in\lbrace 0,\dots,N\rbrace$. The solutions of Eq. $\left(\ref{eq:mastereq}\right)$ and the rates of infection will be denoted by $p_{k_0,k}^{\alpha}$ and $a_k^\alpha$, respectively, when the dependence on additional parameters $\alpha$ needs to be enforced. 

The quality of the surrogate model, i.e., how well it approximates the exact model, depends strongly on the choice of infection rates $a_k$. The way $a_k$ depends on $k$ is determined by the underlying network structure. An analytic formula for $a_k$ is only available for the fully connected network, namely:  $a_k=\tau k(N-k)$, that is the number of S-I links (i.e., links connecting susceptible and infected nodes) in the network multiplied by the per-contact rate of infection $\tau$.

In fact, in a stochastic simulation of the epidemic on a network, the rate of going from $k$ to $k+1$ infected nodes is exactly $\tau \times \# \text{S-I links}$. Hence, during a simulation it makes sense to keep track of the number of infected nodes, the number of S-I links and the time spent in each respective state. Further important observations can be made. The number of S-I links is a random variable and given a fixed number of infected nodes, say $k$, the number of S-I links can take different values. This is simply due to the stochasticity in how the infected nodes are laid out in the network. Thus a plausible choice for the rate $a_k$ may be simply the average of the number of S-I links when there are exactly $k$ infected nodes. However, some states are longer lived than others and this needs to be accounted for. Combining all the above, an empirical average rate of infection emerges, that is
\begin{equation}
	\hat{a}_k=\tau\frac{\sum_{i}it_{i,k}}{\sum_{i} t_{i,k}},\;1\leq k\leq N,
	\label{Eq:average_ak}
\end{equation}
where $t_{i,k}$ is the lifetime of a state with $k$ infected nodes and $i$ S-I links. We will use the notation $\hat{a}_k^{\theta, \tau,\gamma}$ to indicate the resulting estimate given the network class $\theta\in\Theta$: 
$$
\Theta:=\{\mbox{Reg, E-R, B-A}\}.
$$
where we use Regular (Reg), Erd\H{o}s-R\'enyi (E-R) and Bar\'abasi-Albert (B-A) network classes.

\begin{figure}[htbp]
    \centering
        \includegraphics[scale=1]{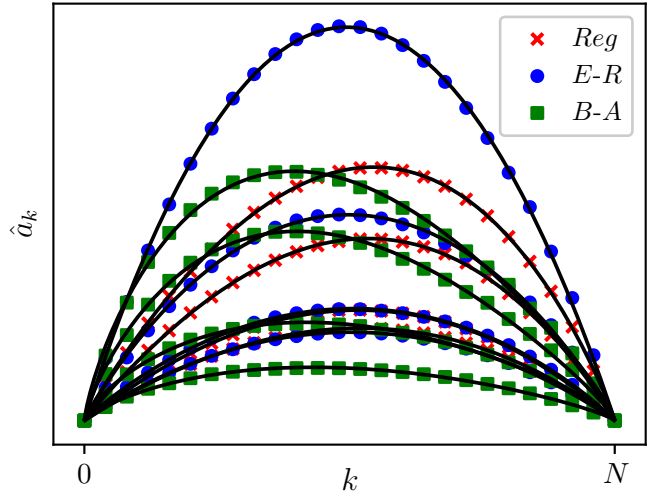}
        \caption{$\hat{a}_k$ curves (markers) along with the best fits from the $C,a,p$ model (plain lines) on $12$ different combinations of network classes and epidemic parameters. Parameters of the simulations considered are, from top to bottom: Reg (crosses), $(\langle k \rangle, \tau, \gamma)=\{(12, 1.43, 5.69)
(6,9.46, 4.23)
(8, 4.47, 8.62)
(13, 1.56, 9.18)\}$; E-R (circles), $(\langle k \rangle, \tau, \gamma)= \{(7,5.96,8,07),(13, 5.8,9.06),(6,3.08,7.61), (16,0.99,8.5)\}$; and B-A (squares), $(\langle k \rangle, \tau, \gamma)=\{(6, 3.09, 7.61)
(8, 5.99, 7.01)
(12, 0.79, 8.96)
(16, 2.18, 5.81)\}$.}
        \label{fig:cap_examples}
\end{figure}

Hence, we can calibrate the infection rates $a_k$ through a statistical analysis based on stochastic simulations of the SIS epidemics on networks. Namely, for a network class (with given average degree) and given disease parameters ($\tau,\gamma$), we run $50$ outbreaks on $50$ different realisations of the network. We keep track of the states that the process visits along with the number of infected nodes, number of S-I links and lifetime of the states. This data feeds into Eq.~\eqref{Eq:average_ak} and leads to the value of $\hat{a}_k$ for all $0\leq k\leq N$. To cover the entire range,  $0\leq k\leq N$, half of the outbreaks are started from $k_0=5$ infected nodes, chosen uniformly at random, and the other from $k_0=N$ infected nodes. The former allows us to explore the curve up to the steady state, while the latter, although an artificial scenario, allows us to explore the curve from the steady state to $N$. Typical $(k, \hat{a}_k^{\theta,\tau,\gamma})$ curves are shown in Fig.~\ref{fig:cap_examples}. In what follows we assume that these rates are `optimal' and that they lead to a surrogate model that agrees well with the exact one.
This choice is motivated by the heuristics presented above which is further validated through extensive numerical simulations for three network classes and a large set of disease parameter values (see Section~\ref{sec:bayesian}).

%
%

\subsection{Three-parameter model of infection rates}
Consistent with results in \cite{Nagy2014}, we notice that, although estimated $\hat{a}_k$ curves are distinct for different network classes, they all share some common features: specifically, they all satisfy $\hat{a}_0=\hat{a}_N=0$ and exhibit a single maximum in $[0,N]$. Perhaps the most important features that change between the three distinct network classes are the flatness and skewness of the $\hat{a}_k$ curves (see Fig.~\ref{fig:cap_examples}). It is clear that high heterogeneity in the degree distribution (i.e. Reg $\rightarrow$ E-R $\rightarrow$ B-A, displaying respectively no $\rightarrow$ medium $\rightarrow$ high heterogenity) increases the left skew. 

The intuitive reason for these differences in the ($k,\hat{a}_k$) curves is that epidemics on such different networks spread with distinct enough characteristics. In scale-free networks for example, the most exposed nodes are the hubs, so they get infected early on. This skews the $(k,\hat{a}_k)$ curve to the left, because once infected these hubs generate a disproportionately large number of S-I links. On the contrary, when all nodes have similar degrees, the $(k,\hat{a}_k)$ curves are more symmetric. Concerning E-R and Reg networks, the most important difference is that the former allows for some degree heterogeneity, whereas the latter does not. Degree heterogeneity plays an important role when it comes to disease transmission so it is no surprise that epidemics on E-R networks can affect a higher proportion of nodes in the initial stage of an outbreak when compared to epidemics on Reg networks~\cite{Kiss2017}.

This suggests that $\hat{a}_k$ curves could be parametrised with a low dimensional model. The departure from the fundamental assumption of homogeneous random mixing in epidemiological and ecological models has led to a myriad of models where bi-linear transmission terms proportional to $\sim I \times S$ or $\sim I \times (N-I)$ have been replaced by non-linear infection terms such as $I^pS^q$~\cite{Liu1986,Hethcote1991,Roy2006}. In particular it is noted that, in the context of classical compartmental and mean-field models, such terms can be inferred from the number of S-I links taken from simulation and that they can lead to more exotic model behaviours. In the same spirit, we put forward the following model for the rates: 

\begin{equation}
\forall k\in\lbrace 0,\dots,N\rbrace,\;a_k:=a_k^{(C,a,p)}=Ck^p \left(N-k \right)^p\left(a\left(k-\frac{N}{2}\right)+N\right),
\label{eq:Capmodel}
\end{equation}
where the three parameters $C$, $a$ and $p$ offer flexibility to adapt to various networks and epidemics of different severity. This choice is well grounded in the literature and is motivated by the heuristic thinking of how the epidemic unfolds on the network. The parameter $C>0$ gives a general scaling, dealing with different infection intensities, $a\in[-2,2]$ helps to shift the peak from the centre (e.g. $a<0$ shifts the peak to the left), and $p>0$ allows for different flatnesses (smaller $p$ values leading to flatter curves). Immediately, one can note that this model fulfils a number of desirable properties: (a) it is low dimensional/parsimonious, (b) the model satisfies $a_0=a_N=0$ by construction, (c) it includes the complete network when $a_k=\tau k(N-k)$ and finally, (d) it has a single maximum within $[0,N]$.

The $C,a,p$ values are obtained using a non-linear least-square fit (using a particle swarm algorithm \cite{Particleopt}):
\begin{equation}
e(C,a,p;\mathcal{S})=\sum_{k,\;\sum_{i}t_{i,k}>0}\left(a_k^{(C,a,p)}-\hat{a}_k\right)^2.
\label{eq:ls}
\end{equation}
Fig.~\ref{fig:cap_examples} showcases the flexibility of the model in fitting $\hat{a}_k$ curves coming from different network classes and confirms our observations about the rates being more left-skewed with increasing heterogeneity in node degree.

In the same figure, curves based on the ($C, a, p$) model are compared to the $(k,\hat{a}_k)$ curves. Systematic numerical investigations (not all plots shown) demonstrate that the proposed parsimonious three-parameter model fits the $(k,\hat{a}_k)$ curves well for all considered network classes, particularly Reg and E-R. For B-A networks small discrepancies between the ($k,\hat{a}_k$) curves and the ($C,a,p$) model are possible.
%
%

\subsection{Dataset}
Proving that the behaviour of the exact system of $2^N$ equation is well approximated by our proposed system of $(N+1)$ ordinary differential equations~\eqref{eq:mastereq} is still an open question. Therefore, the validations that we provide in this paper are entirely based on extensive numerical simulations. Here, we discuss briefly the synthetic dataset $\mathcal{S}$ underpinning those numerical validations.  
For each network class, we varied the average degree ($5\leq\langle k \rangle < 20$). This covers a large number of scenarios and the networks remain relatively sparse.
Regarding the epidemic parameters, we varied the infection and recovery rates ($(\tau, \gamma) \in (0,10] \times (0,10]$). Values for the rates were chosen via Latin hypercube sampling~\cite{Conover1979}.
By doing so, we could observe many unique scenarios, providing a solid base upon which to test our methods. 

However, there may be situations where the epidemic does not spread. Indeed, the behaviour of an epidemic is determined by the characteristics of both the network class and epidemic dynamics. The former includes quantities such as the average degree and higher-order moments, the latter includes per-link infection and recovery rates. All of this is captured by the reproduction number~\cite{Kiss2017}, $R_0$, which is the number of secondary infections caused by a typical infectious individual introduced into a fully susceptible population:
\begin{equation}
R_0 = \frac{\tau}{\gamma+\tau} \frac{\braket{k^2 - k}}{\braket{k}}.
\label{epidemicratio}
\end{equation}
If $R_0\leq 1$ the infection will die out. However, if $R_0>1$, then an outbreak is expected. Since $R_0$ depends directly on the sampled network and disease parameters, we accepted only situations where $1<R_0\leq 10$. This led to $360$ valid choices ($\text{network class}, \langle k \rangle, \tau, \gamma$), with 120 per network class. For all the 360 scenarios, 
data from the simulations were used to determine network class- and disease-parameter specific infection rates $\hat{a}_k^{\theta, \tau,\gamma}$ and the corresponding ($C, a, p$) models.

\subsection{Numerical validation of the forward model}
\label{subsec:num_val_forward_model}
\begin{figure}[htbp]
	\centering
	\begin{subfigure}[b]{0.32\textwidth}
		\centering
		\includegraphics[scale=1]{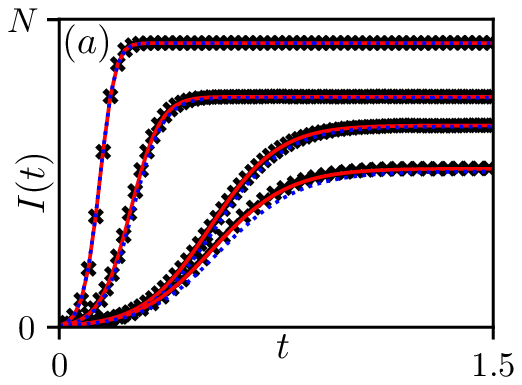}
	\end{subfigure}%
	~ 
	\centering
	\begin{subfigure}[b]{0.32\textwidth}
		\centering
		\includegraphics[scale=1]{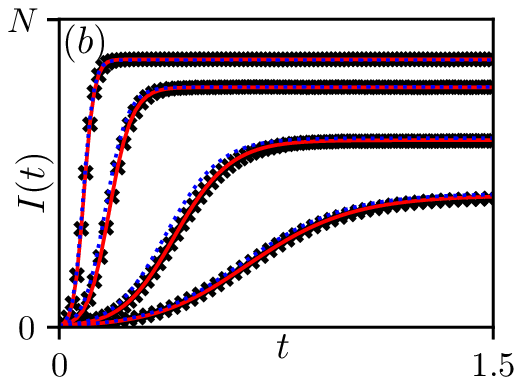}
	\end{subfigure}%
	~    
	\centering
	\begin{subfigure}[b]{0.32\textwidth}
		\centering
		\includegraphics[scale=1]{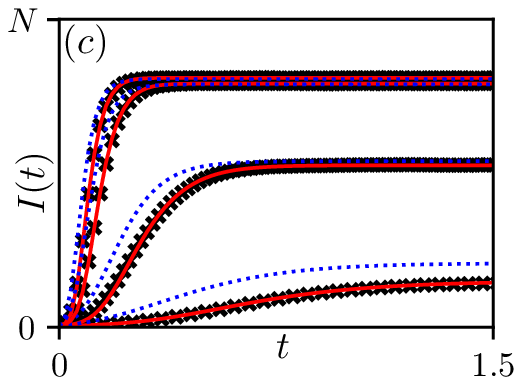}
	\end{subfigure}%
	\caption{Average number of infected nodes from simulations (markers) and the numerical solutions of system (\ref{eq:mastereq}) with rates $a_k$ given either by the raw data $\hat{a}_k$ (continuous curve) or by the ($C, a, p$) model (dotted curves), with initial condition $k_0=5$. Three network classes are reported, each with $N=1000$ nodes, from left to right, ordered by increasing heterogeneity, from Reg (a) and E-R (b) to B-A (c) networks. Networks and epidemic parameters are the same as in Fig.~\ref{fig:cap_examples}.}
	\label{fig:MastereqvsEpid}
\end{figure}

To validate our claim that the BD process is a good approximation of the true epidemic behaviour, we numerically integrated the master equation~(\ref{eq:mastereq}) with rates $a_k = \hat{a}_k$ and $c_k = \gamma k$, where $\hat{a}_k$ are the estimated rates via Eq.~(\ref{Eq:average_ak}), for all 360 scenarios in $\mathcal{S}$. The master equation was also numerically integrated with rates given by the ($C, a, p$) model. The expected number of infected nodes from the numerical solution of both master equations was then compared to the average number of infected nodes based on simulations. Four representative examples of epidemics for each network class are shown in Fig.~\ref{fig:MastereqvsEpid}. 
For the vast majority of the tested cases (not all shown), the agreement between simulation and the $(C,a,p)$ model is good to excellent. 
It is worth noting that, in the case of B-A networks there are a few parameter combinations where the agreement between the master equation with the rates given by the ($C,a,p$) model and simulation results is poorer, see Fig.~\ref{fig:MastereqvsEpid}(c). This is despite the seemingly small discrepancy between ($k,\hat{a}_k$) curve and the corresponding ($C,a,p$) model (not shown). However, the master equation with the $\hat{a}_k$-rates still leads to good agreement with simulations as shown in Fig.~\ref{fig:MastereqvsEpid}(c) (markers versus continuous line).
Even so, it is reassuring to see that even when the agreement between the master equation with the ($C, a, p$) model breaks down, the agreement with the $\hat{a}_k$ holds. In~\cite{Nagy2014}, a similar surrogate model was used and the authors obtained good agreement between the BD model and simulations for an even wider range of network classes. This gives us confidence that the surrogate model is a viable model.


\section{Bayesian inference of network class from single epidemics}
\label{sec:bayesian}

In the framework presented so far, we proposed a surrogate model which approximates the evolution of the total number of infected nodes in a SIS epidemic on a network. The rates of infection in this forward model (i.e. $\hat{a}_k$) are parametrised by a three-parameter model ($C,a,p$) as detailed in equation~\eqref{eq:Capmodel}. Early investigation shows that the ($k,\hat{a}_k$) curves (thus the associated $C,a,p$ triple) are distinct across the three different network classes that we considered. Hence, one may expect that discrete observations taken from a single epidemic spreading on a unknown network carry sufficient information to identify its most likely class.

\begin{figure}[h!]
	\centering
	\begin{subfigure}[b]{0.315\textwidth}
		\centering
		\includegraphics[scale=1,trim={0cm 0.5cm 0cm 0cm}]{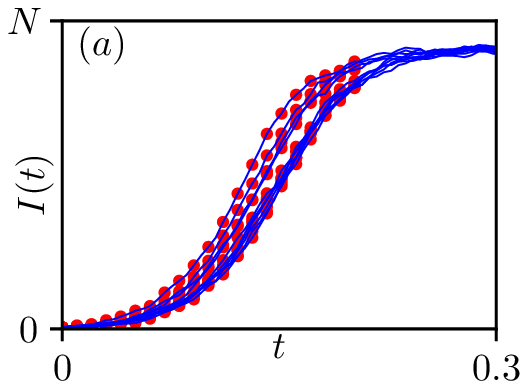}
		\label{sim1}
	\end{subfigure}
	~
	\begin{subfigure}[b]{0.315\textwidth}
		\centering
		\includegraphics[scale=1,trim={0cm 0.5cm 0cm 0cm}]{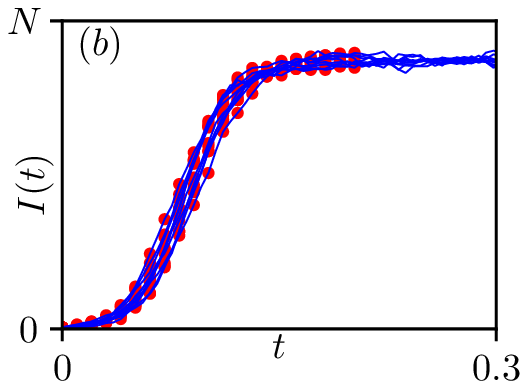}
		\label{sim2}
	\end{subfigure}
	~
	\begin{subfigure}[b]{0.315\textwidth}
		\centering
		\includegraphics[scale=1,trim={0cm 0.5cm 0cm 0cm}]{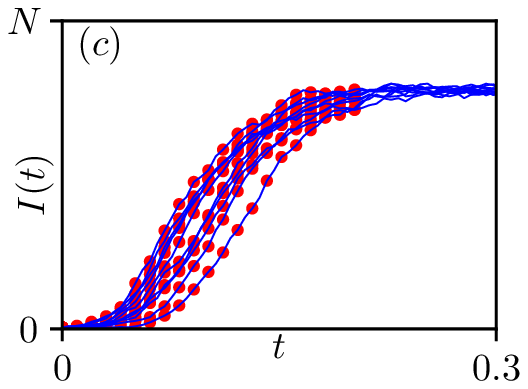}
		\label{sim3}
	\end{subfigure}
	\caption{10 independent realisations of datasets for 3 examples on different network classes. Continuous curves represents the evolution of infectious counts and red dots are the observations $y$. Network and epidemic parameters for each panel are, from left to right, Regular  with $\langle k \rangle=17$, $\tau=2.62$, $\gamma=4.03$, \ER with $\langle k \rangle=13$, $\tau=5.80$, $\gamma=9.06$, and Barab\'asi-Albert with $\langle k \rangle=6$, $\tau=8.16$, $\gamma=8.23$.}
	\label{fig:sims}
\end{figure}

To be more precise, we consider a population-level dataset $y=(k_1,\dots,k_n)$ where $k_j\in\lbrace 0,\dots,N\rbrace$ for any $j=1,\dots,n$ is the number of infected nodes in the network at time $t_j\in[0,T]$, and we define the vector $s=(t_1,\dots,t_n)$. Our objective is to predict the class $\theta\in\Theta$ of the underlying network from $y$. Figure \ref{fig:sims} illustrates 10 distinct data sets for each of the three network classes. These data are obtained directly from Gillespie simulations~\cite{Gillespie1976, Gillespie1977} of the SIS epidemic on the respective networks. Observations are taken at regular times from the start of the epidemic to the point where the quasi steady-state is approached.

For each value of $\theta$ (that is a network class), we build a distribution $\pi_{0,\theta}$ over the parameters $C,a,p$ based on offline simulations of SIS epidemics for a range of networks in each given class $\theta$ (see Section \ref{subsec:prior}). By looking at the outcomes of our simulations, we observe that, for our chosen set of candidate classes $\Theta$, the distributions $\pi_{0,\theta}(C,a,p)$, $\theta\in\Theta$, cluster in distinct regions of the $(C,a,p)$ parameter space. This is necessary for the inference to work, and it  contributes to the validation of our model of choice for the rates $\hat{a}_k$. Assuming a non-informative uniform prior for $\theta$, we derive a prior distribution $\pi_0(C,a,p,\theta)$ in the form of a mixture:
$$
\pi_0(C,a,p,\theta)=\frac{1}{3}\pi_{0,\theta}(C,a,p).
$$
Our objective is the prediction of the underlying network $\theta$ given the data $(y,s)$, which will be done using the posterior distribution $\pi(\theta\vert y,s)$ obtained by Bayes' rule:
\begin{equation}
\begin{split}
\pi(\theta|y,s)&=\int\pi(C,a,p,\theta\vert y,s)dCdadp\\
&\propto\int\mathcal{L}^{C,a,p}(y,s)\pi_{0}(C,a,p,\theta)dCdadp,\\
&\propto\int\mathcal{L}^{C,a,p}(y,s)\pi_{0,\theta}(C,a,p)dCdadp,
\end{split}
\label{Eq:Marginal}
\end{equation}
where, given $C,a,p$, the likelihood $\mathcal{L}^{C,a,p}$ can be expressed in terms of the solution operator of the forward model discussed above (see Section \ref{sec:forward_model}). This Bayesian classification methodology is also known as model selection, where the model is a particular class of networks. Once we have computed the posterior distribution $\pi(\theta\vert y,s)$ (see Section~\ref{subsec:computations}), we simply pick the most likely underlying network class (Maximum a posteriori estimator for $\theta$ given the data $(y,s)$).

\subsection{Prior distributions for each network class}
\label{subsec:prior}

\begin{figure}
	\centering
	\includegraphics[scale=1,trim={2cm 1cm 0cm 3cm}]{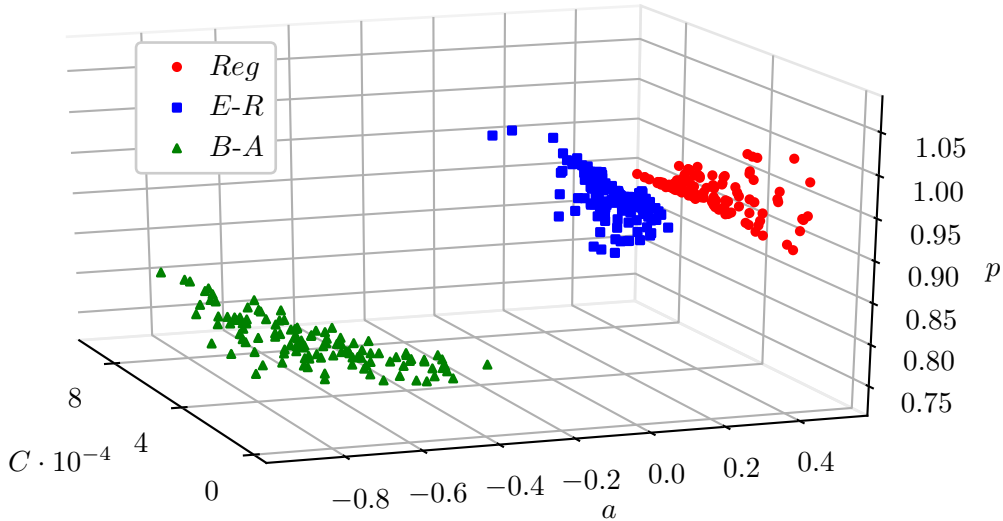}
	\caption{Estimated $C,a,p$ values from the dataset $\mathcal{S}$ ($360$ points in total, each coming from a unique combination of (network class, $\langle k \rangle$, $\tau$, $\gamma$)). From left to right, we observe three distinct regions corresponding to Barab\'asi-Albert (triangles),\ER (squares) and Regular networks (circles) networks.}
	\label{fig:3dand2dclassification}
\end{figure}

In this work, we consider prior distributions for each network class as a different density $\pi_{0,\theta}$ over the $C,a,p$ space. To do this, we use the very same dataset that was used for numerical validation (see Section~\ref{sec:forward_model}). Given the $C,a,p$ values of each network class (see Fig.~\ref{fig:3dand2dclassification}), we  choose $100$ triples to estimate a probability distribution and leave $20$ for testing. The ($C,a,p$) values associated with the training scenarios are used to infer three Gaussian kernel density estimators \cite{scikit-learn} to be used as prior distributions. The bandwidth of these estimators is set by $10$-fold cross-validation.

\subsection{Numerical method for posterior marginals computations}
\label{subsec:computations}

Finally, to predict the underlying network class given a dataset $(y,s)$, we need to compute the three marginals given in equation~\eqref{Eq:Marginal} (one per network class) and this is done by Monte-Carlo estimation.
As already mentioned, the likelihood $\mathcal{L}^{C,a,p}(y,s)$ can be obtained using the forward operator associated with equation~\eqref{eq:mastereq}. Indeed, given a ($C,a,p$)-triple, the likelihood of a dataset $(y,s)$ is given by:
\begin{equation*}
	\mathcal{L}^{C,a,p}(y,s)=\prod_{i=1}^{n-1}p_{k_i,k_{i+1}}^{C,a,p}(t_{i+1}-t_i),
\end{equation*}
using the fact that the BD process is time homogeneous. We choose to compute each term $p_{k_i,k_{i+1}}^{C,a,p}(t_{i+1}-t_i)$ where $1\leq i\leq n-1$ using the algorithm introduced in \cite{Crawford14}, allowing BD transition probabilities to be computed individually. This represents a significant reduction in computational time, when compared to matrix exponential since we are working with a network of size $N=1000$ nodes.

Once we have an efficient numerical method to compute the likelihood, we use the corrected Arithmetic Mean estimator, recently introduced in \cite{Pajor2017} for the Monte-Carlo estimation of all marginals. Let $A$ be a given subset of the $(C,a,p)$ space, then it follows that:
\begin{equation}
\begin{split}
\int\mathcal{L}^{C,a,p}(y,s)\pi_{0,\theta}(C,a,p)dCdadp=\frac{\pi_{\theta,0}(A)}{\pi_{\theta\vert (y,s)}(A)}\int_A\mathcal{L}^{C,a,p}(y,s)\pi_{A,\theta}(C,a,p)dCdadp,
\end{split}
\label{Eq:aAME}
\end{equation}
where $\pi_{A,\theta}$ is the prior density of network class $\theta$, conditional on $\theta\in A$. Each marginal is then estimated using the following procedure:
\begin{enumerate}
	\item Find
	\begin{equation*}
		(C^*,a^*,p^*)=\arg\max_{C,a,p}\left(\log\mathcal{L}^{C,a,p}(y,s)+\log\pi_{\theta,0}(C,a,p)\right).
	\end{equation*}
	This is done via a combination of global/local optimisation routines.
	\item Sample the distribution $\pi_{\theta}(C,a,p\vert y,s)$ using a Random-Walk Metropolis-Hastings algorithm starting from $(C^*,a^*,p^*)$ and denote the samples by $(C_i,a_i,p_i)_{1\leq i\leq K}$ with $K=500$.
	\item Let $H$ be the Fisher information evaluated at $(C^*,a^*,p^*)$ and let $d(C,a,p)$ be defined as
	\begin{equation*}
		d(C,a,p):=\left\langle (C_i,a_i,p_i)-(C^*,a^*,p^*), H\left[(C_i,a_i,p_i)-(C^*,a^*,p^*)\right]\right\rangle.
	\end{equation*}
	We then take $A:=\left\lbrace (C,a,p)|d(C,a,p)\leq r\right\rbrace$ where $r=\max_{1\leq i\leq K}d(C_i,a_i,p_i)$. This choice was already suggested in \cite{Pajor2017}. In particular, it leads to $\pi_{\theta\vert (y,s)}(A)\approx1$, which simplifies the right-hand-side of equation~\eqref{Eq:aAME}.
	\item Use a Gaussian distribution $\mathcal{N}\left((C^*,a^*,p^*),H^{-1}\right)$ to estimate both $\pi_{\theta,0}(A)$ and the integral term on the right-hand-side of equation~ \eqref{Eq:aAME} by importance sampling.
\end{enumerate}
Our complete Python implementation of this routine is available online \footnote{\url{https://github.com/BayIAnet/NetworkInferenceFromPopulationLevelData}}.

\subsection{Network class inference}

In this section, we provide numerical results assessing the overall quality and applicability of our approach. We start by inferring networks from a testing dataset, where all data are simulated from either Regular, E-R or B-A networks, see Section~\ref{subsec:inf_test_scen}. We then consider networks outside of our framework, namely synthetic networks with negative binomial degree distributions (Section~\ref{subsec:synthetic}) and real-world networks (Section~\ref{subsec:realnetworks}). In all cases, we provide posterior probabilities for each network class across independent repetitions of the datasets to quantify uncertainty.

\subsubsection{Inference based on the testing set}
\label{subsec:inf_test_scen}
During the construction of the prior, we deliberately set aside $60$ estimated $(C,a,p)$ values to build a test set (20 per network class taken at random), meaning that they were not used in the calibration of the prior. In this section, we use this set to check if we can infer the known underlying network class.

The inference was performed as follows. For each of the ($C,a,p$) parameters in the testing set, we used the known underlying network and disease parameters ($\text{network class}$, $\langle k \rangle$, $\tau$, $\gamma$) to simulate a dataset $(y,s)$ with Gillespie's algorithm. We only generated a single network from the appropriate class and simulated a single epidemic. However, we generated 10 independent datasets, as shown in Fig.~\ref{fig:sims}, and ran our inference model on each of them separately. The second step was to compute the 3 posterior probabilities corresponding to the different network classes, as detailed earlier. We thus obtained 3 posterior probabilities for each of the $60$ elements in our test set and predicted the most likely underlying network class. To assess the uncertainty due to data sampling, we considered the results across all the independent datasets.

The quality of the inference is shown by the confusion matrix (Table~\ref{table:conf}), which provides the averaged posterior probabilities along with their standard deviation for each of the possible outcomes. The level of accuracy achieved in our tests is remarkable, with a score as high as $95\%$ for Barab\'asi-Albert, and a minimum of $79\%$ for Erd\H{o}s-R\'enyi. This also shows that there can be a moderate confusion between the Regular and Erd\H{o}s-R\'enyi network classes, as their characteristics are quite similar w.r.t. ($C,a,p$) values (see Fig.~\ref{fig:3dand2dclassification}) whereas Barab\'asi-Albert is rarely miss-classified. Further, the standard deviations show that these scores are stable across different data realisations, suggesting that our approach is consistent.

\begin{table}[h!]
	\centering
	\begin{tabular}{cccc}
		\hline 
		True/Predicted & Regular & Erd\H{o}s-R\'enyi & Barab\'asi-Albert \\ 
		\hline
		Regular & $85.5\%$ ($7.9\%$) & $14.5\%$ ($7.9\%$) & $0.0\%$ ($0.0\%$)\\ 
		Erd\H{o}s-R\'enyi & $21.5\%$ ($10.7\%$) & $78.5\%$ ($10.7\%$) & $0.0\%$ ($0.0\%$) \\ 
		Barab\'asi-Albert & $0.0\%$ ($0.0\%$) & $5.0\%$ ($5.0\%$) & $95.0\%$ ($5.0\%$) \\
		\hline 
	\end{tabular}
	\caption{Averaged confusion matrix based on the test dataset (standard deviation is brackets).}
	\label{table:conf}
\end{table}

\begin{figure}[h!]
	\centering
	\includegraphics[width=6in,height=3in]{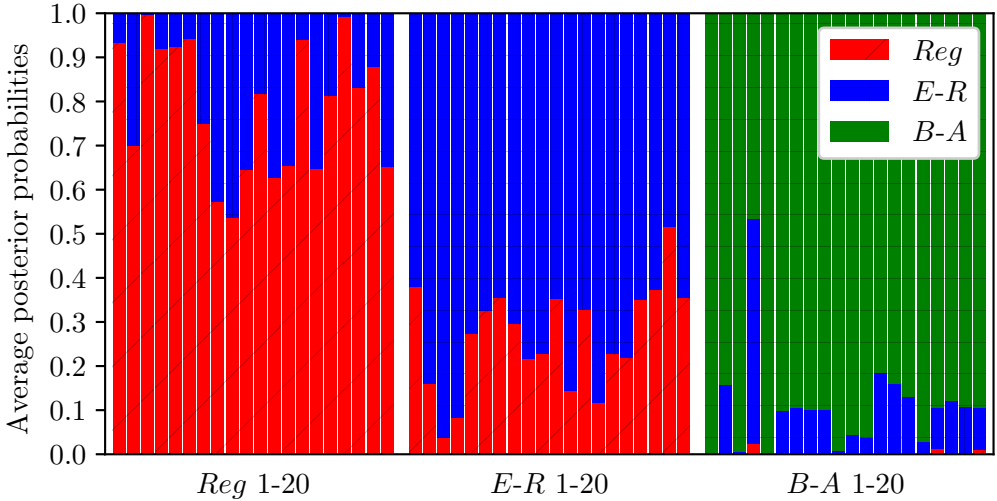}
	\caption{Average posterior probability over the $60$ tests ($20$ per network class and $10$ realisations).}
	\label{fig:classifier}
\end{figure}

To get a more precise description of the classification results, we computed the average posterior probability for each of the $60$ test elements, see Fig.~\ref{fig:classifier}. This revealed that the average posterior probability varies within each of the network class, probably due to differences in network or disease parameters. 
In some sense, this shows that for some network and disease parameters, the similarity between Regular and Erd\H{o}s-R\'enyi is significant. For example, when the epidemic spreads fast and infects many nodes early on, the structure of the network is less important as the infection will be transmitted on. This means that the average degree is more important than the degree distribution. Nevertheless, our inference methodology returns a good classification in most cases. In fact, these tests show that our approach can successfully recover the network class from as little as 21 observations of a single epidemic.

\begin{figure}[h!]
	\centering
	\includegraphics[width=5in,height=4in]{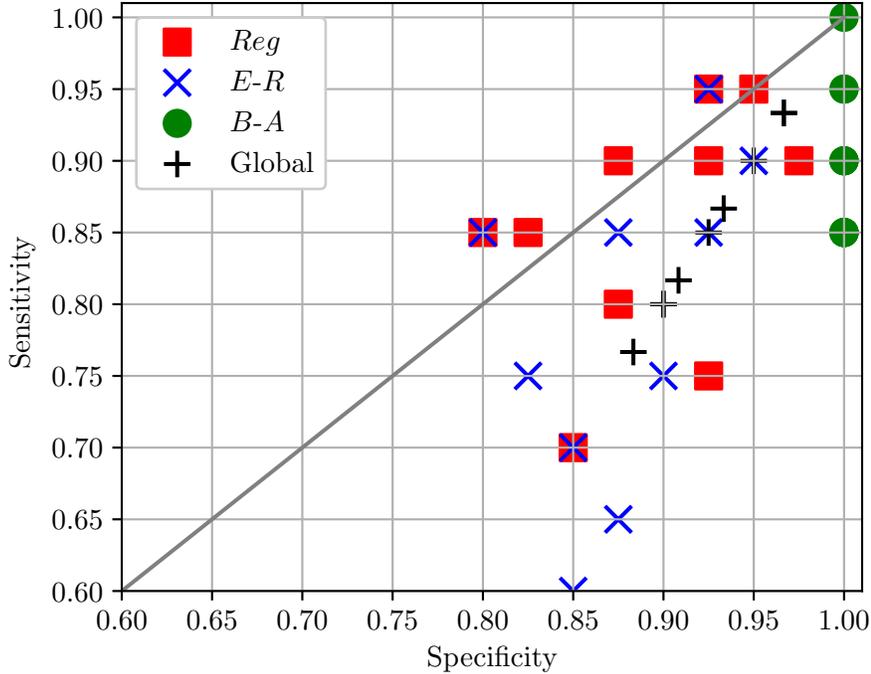}
	\caption{Specificity and sensitivity of the $10$ independent classification at global and per-network levels.}
	\label{fig:sensi}
\end{figure}

Finally, we detail specificity and sensitivity for the $10$ repetitions of the classification, offering per network class and global statistics in Fig.~\ref{fig:sensi}. We note that each marker has 10 occurrences but in some cases these are superimposed. Here again, one can see the stability and high efficiency of our approach for Barab\'asi-Albert, with more confusion for Erd\H{o}s-R\'enyi.

\subsubsection{Inference of synthetic networks}
\label{subsec:synthetic}

We have shown that our methodology performs well when applied to the data generated on the networks that it was trained on. In this Section, we consider alternative network types for two reasons: (a) to stress-test the classification by using networks whose degree distributions do not come from the models used to build priors, and (b) to study the extent to which it can distinguish between different levels of heterogeneity in degree distribution.

\begin{figure}[h!]
	\centering
	\begin{subfigure}[b]{0.46\textwidth}
		\centering
		\includegraphics[scale=1]{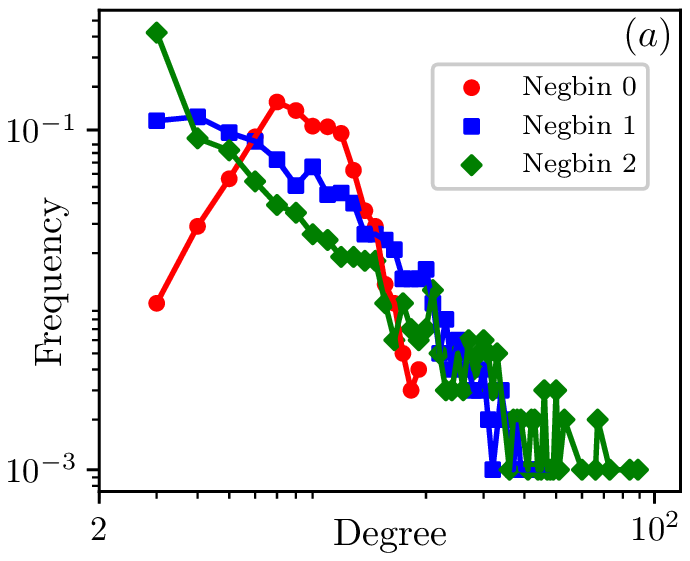}
	\end{subfigure}
	~
	\begin{subfigure}[b]{0.46\textwidth}
		\centering
		\includegraphics[scale=1]{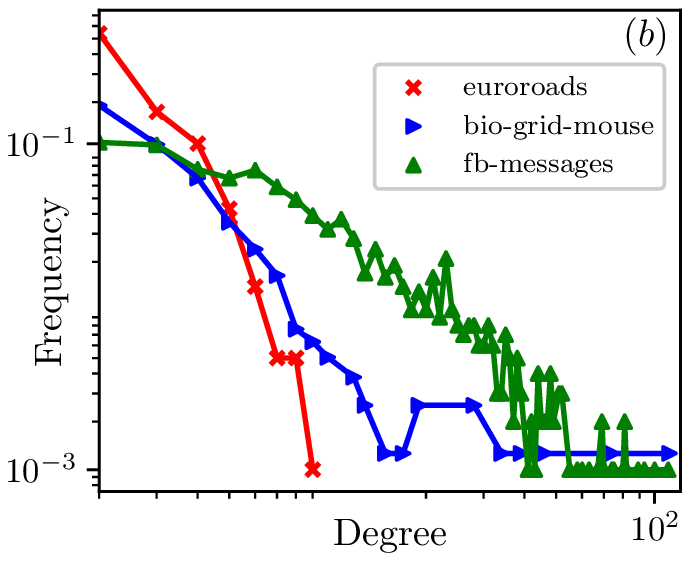}
	\end{subfigure}
	\caption{Degree distributions, ordered by variance, of three single negative binomials (a) following equation (\ref{eq:negbin}) and of the three real networks (b) used for the stress test. For (a), the average degree is $\langle k \rangle = 6$ for all networks. From low to high variance we have $\sigma =  8$ (Negbin 0), $\sigma =  40$ (Negbin 1), $\sigma =  120$ (Negbin 2). The values of $(p,n)$ are $(0.23,20),(0.85,1),(0.95,0.3)$, respectively. For (b) the basic metrics of these networks are $\{\langle k \rangle, \sigma^2 , \text{Assortativity}, \text{Clustering}\} =
\{2.53, 5.24, 0.102, 0.02\} , \{2.77, 40, -0.21, 0.04\} , \{12.30, 268.90, -0.08, 0.09\}$, respectively.}
	\label{fig:negbindegreedist}
\end{figure}

\begin{figure}[h!]
	\centering
	\includegraphics[width=6in,height=3in]{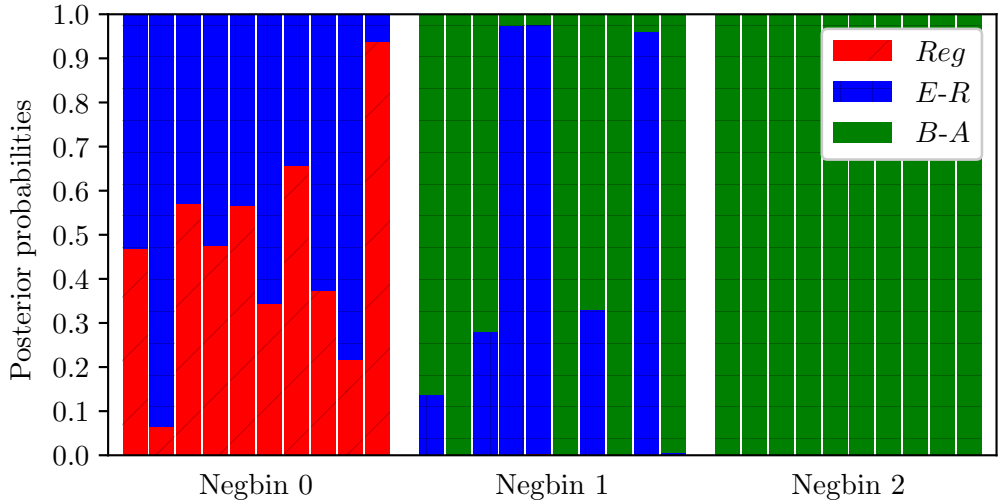}
	\caption{Posterior probabilities for the 10 repetitions on each synthetic network.}
	\label{fig:synthetic_classifier}
\end{figure}

To do this, we generated three synthetic networks using the configuration model~\cite{Configurationmodel} and a negative binomial degree distribution with parameters $(p,n)$:
\begin{equation}
\forall k\in\lbrace 0,n\rbrace,\;P(k) = { k+n-1 \choose k }p^k(1-p)^{n-k},
\label{eq:negbin}
\end{equation}
where $p$ is the probability of success and $n$ the number of failures. This choice is motivated by both the simplicity and the flexibility of this distribution. The average degree in all three networks was identical (i.e. fixed at $\langle k \rangle = 6$) but with different levels of heterogeneity depending on the variance, see Fig.~\ref{fig:negbindegreedist}(a). To avoid the possibility of having disconnected components, the degree distributions were shifted so that the minimum was greater or equal to $3$. Here, the degree distributions were chosen to exhibit different levels of heterogeneity, from low to a level comparable to those achieved in B-A networks. We then ran $10$ independent epidemics with parameters $\gamma = 1$ and $\tau = 0.5$, starting from $5$ infected nodes. As in Section~\ref{subsec:inf_test_scen}, the inference was based on a dataset with $21$ equally-spaced observations of the number of infected nodes. The results are shown in Fig.~\ref{fig:synthetic_classifier}, and confirm that our inference scheme is able to distinguish between networks with high/low levels of degree heterogeneity. In particular, by looking at Fig.~\ref{fig:negbindegreedist}{(a)} it is reasonable to expect that the first and third networks are going to be classified as E-R and B-A networks, respectively. Indeed, Fig.~\ref{fig:synthetic_classifier} shows that the first network in Fig.~\ref{fig:negbindegreedist}(a) is identified as E-R 80\% of the time, whereas the third network in Fig.~\ref{fig:negbindegreedist}(a) is correctly classified as B-A for every single epidemic realisation.


When the degree distribution of the test network is such that its variance falls between typical variances observed in E-R and B-A networks (see the second network in Fig.~\ref{fig:negbindegreedist}-a) our results are more sensitive to the individual realisation of the epidemic. However, even in this case, the network is identified to the closest type in terms of degree distribution. Moreover, heuristically at least, the B-A network seems to be favoured, which seems reasonable upon inspecting the degree distribution of the test network.

\subsubsection{Inference of real-world networks}
\label{subsec:realnetworks}

Finally, the last test we conducted was based on real world networks, which can exhibit higher-order structure beyond degree heterogeneity. We chose three real networks: the first is labelled \textrm{euroroads} and is part of the KONECT collection \cite{Konectdataset}, the second and third, \textrm{bio-grid-mouse} and \textrm{fb-messages}, are part of the network data repository Networkrepository \cite{nr}. The euroroads is an infrastructure network, bio-grid-mouse is a protein-protein network whilst fb-messages is based on the interactions of an online community of students at University of California. In Fig.~\ref{fig:negbindegreedist}(b) the degree distributions of these networks are shown. To keep the number of nodes equal to $N=1000$, we only considered the largest connected component, and then, where necessary, removed peripheral, low-degree nodes such that the resulting network was still connected.

\begin{figure}[h!]
	\centering
	\includegraphics[width=6in,height=3in]{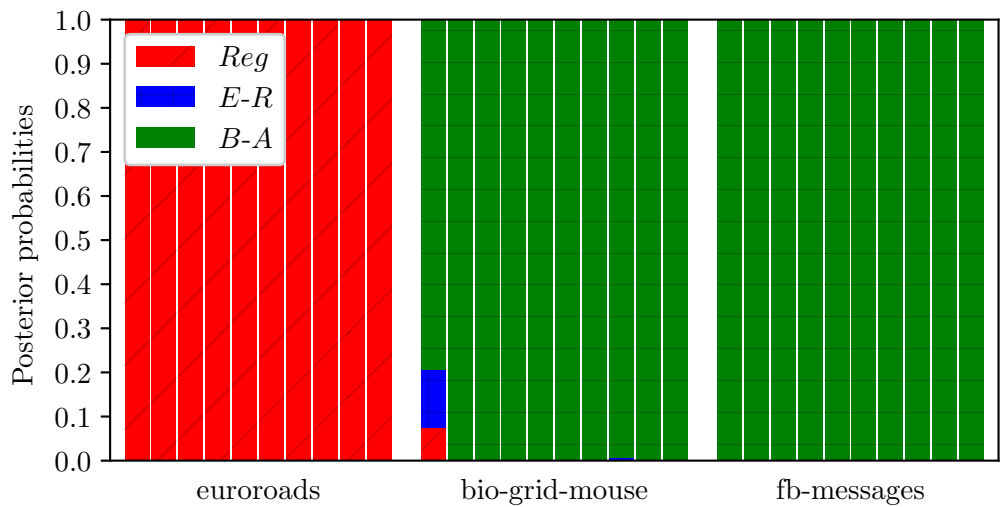}
	\caption{Posterior probabilities for the 10 repetitions on each real-world network.}
	\label{fig:realnet_degree_classifier}
\end{figure}

In line with Section~\ref{subsec:synthetic}, we fixed $\gamma = 1$, and ran $10$ distinct epidemics on each network in order to generate data for the inference. Values for the infection parameter were $\tau = \{1.5, 2.5, 0.4\}$ for euroroads, bio-grid-mouse and fb-messages, respectively. The posterior probabilities obtained from our approach are reported in~Fig.~\ref{fig:realnet_degree_classifier} and are in line with our expectations based on the inspection of the respective degree distributions: the infrastructure network is very homogeneous, whilst the other two are scale-free, and hence  correctly classified as B-A.

\section{Discussion}
\label{sec:discussion}
In this paper, we proposed a new inference scheme that uses population-level incidence data at discrete regular times to infer the most likely network class over which the epidemic has initially spread. This is a challenging task because the exact epidemic model on a given network is forbiddingly high-dimensional meaning that even a numerical solution is out of reach. The key to carry out the inference is the approximation of the exact epidemic model by a BD process, whose rates not only encode the structure of the networks but also allow us to distinguish between the different network classes through a parsimonious three-parameter model. Whilst we have successfully numerically validated this surrogate model over a number of network classes and different values of disease parameters, with further evidence in~\cite{Nagy2014}, a mathematical characterisation of the relation between the exact and this surrogate model remains an open problem.

Our analysis has focused on three well-known classes of random networks: Regular, Erd\H{o}s-R\'enyi and Barab\'asi-Albert. For each network class, the rates of infection in the corresponding BD approximation was obtained by using the time-weighted mean of S-I link counts. Despite these rates being network class-dependent, they all share some common features. This in turn allowed us to propose a parsimonious three-parameter model ($C,a,p$) that works across all network classes and, at the same time, can capture the differences in the rates of the approximating BD process.
In addition to being robust to different values of $\tau$, $\gamma$ and average degree, these parameters exhibit a clear distinction between the three different network classes when plotted in the 3-dimensional ($C, a, p$) space. This knowledge is then encoded into prior distributions, constructed using kernel density estimators over the ($C, a, p$) space. Our Bayesian model selection procedure then consists in the numerical estimation of the relative marginal probabilities. Our results show that the inference scheme has good specificity and sensitivity, despite the simplicity of the model. 
\begin{figure}[h!]
	\centering
	\includegraphics[scale=1]{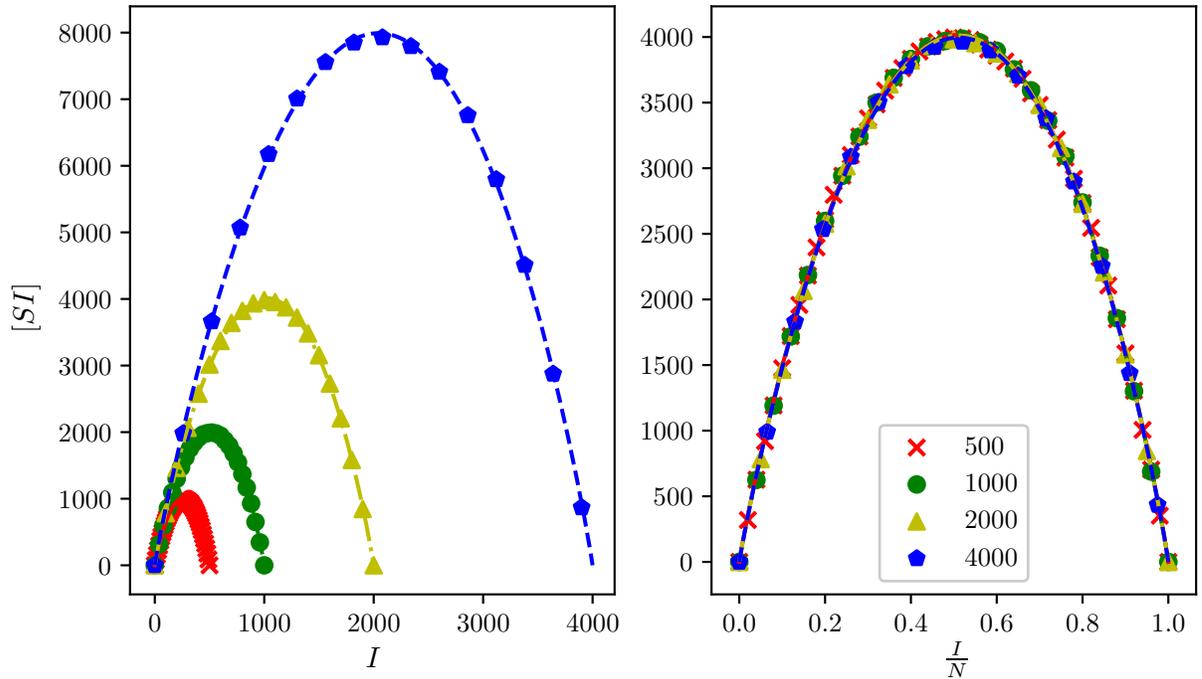}
	\caption{(Left) ($k,\hat{a}_k$) curves based on \ER networks with $\langle k \rangle = 5$, $\tau = 1.793$, $\gamma = 3.785$ and $N= 500, 1000, 2000, 4000$. (Right) Scaled  $(k, \hat{a}_k)$ curves relative to the $N=2000$ case. Scaled version are obtained by plotting ($k/N, \frac{2000}{N}\hat{a}_k$).}
	\label{fig:scaling}
\end{figure}
These encouraging results lead to a number of questions and remarks. First of all, our choice of classes of random networks means that the main feature of the networks is their degree heterogeneity. We have yet to consider more complex networks, such as those exhibiting clustering or community structure. This would certainly lead to ($k,\hat{a}_k$) curves of different shapes, potentially having other features such as multiple peaks for networks with multiple communities, and thus requiring either a more sophisticated or non-parametric model. Nevertheless, considering epidemics in terms of an approximate BD process appears to be a powerful approach if a tractable likelihood is desired. Moreover, once the most likely network class has been identified, one could continue and estimate  $\tau$, $\gamma$ and the average degree.

We have used a fixed number of nodes ($N=1000$) in all our numerical experiments. We do not expect major changes when the number of nodes is different. Preliminary numerical tests, see Fig.~\ref{fig:scaling}, suggest that there is a good degree of universality such that the ($k,\hat{a}_k$) curves only differ by a scaling factor when the number of nodes changes, all other parameters being fixed. In this respect, our methodology could easily be adapted by directly considering the scaled epidemic (on $[0,1]$) and repeating our tests for different values of $N$. 
Fortunately, our numerical method \cite{Crawford12} scales well with $N$, since the transition probabilities in the likelihood are computed individually (with deeper continued fractions). The question of the limiting behaviour in the limit of large $N$ can also be further investigated.

So far, we have used discrete data taken on a regular time grid covering the epidemic from its early stage (a few infectious nodes) up to its steady state. Increasing the frequency of data or restricting data to the very beginning of the epidemic are of significant practical interest. In the former case, one expects the discrete likelihood to converge to the simpler continuous one, enabling faster and easier analysis. In the latter case, it would lead to a model that does not require describing the whole epidemic as we currently do. Focusing on the initial stages of the epidemic, the most critical period in many cases, and upon solving a potential un-identifiability problem, such an approach could have an important real-world impact, making it possible to predict and control more accurately yet-to-be epidemics.


Finally, the proposed inference scheme could be improved by using more sophisticated models for the infection rates and by learning a larger number of different network classes, leading to a wide portfolio of data which can then be used for estimation. Of course, there is a trade-off in terms of what we can infer about networks using population-level discrete data. We cannot infer individual links for example but this is to be expected since the data we use for inference is not at the link- or node-level. Nevertheless, we believe that our approach could have practical implications, as the inference scheme is based on the kind of data that is most likely to be available in real-world scenarios (e.g. the number of infected people every day or week). Where such data is available but little is know about the contact pattern, our inference scheme may be able to provide some high-level information about the properties of the network which in turn could be exploited in the planning or implementation of control, in particular during the early stages of an epidemic.

\section{Acknowledgments}
All authors acknowledge support from the Leverhulme Trust for the Research Project Grant RPG-2017-370.
\clearpage

\bibliography{main}
\bibliographystyle{plain}
\end{document}